\documentclass[12pt,a4paper]{article}

\usepackage{amsmath}
\usepackage{amsfonts}
\usepackage{amssymb}
\usepackage{latexsym}

\makeatletter
\@addtoreset{equation}{section}
\makeatother

\newcommand{\be}{\begin{equation}}
\newcommand{\ee}{\end{equation}}

\newcommand{\eq}[1]{(\ref{#1})}
\newcommand{\fr}{\frac}
\newcommand{\tf}{\tfrac}

\newcommand{\df}{\textrm{d}}
\newcommand{\expe}[1]{\textrm{e}^{#1}}

\newcommand{\sr}{\sqrt}

\newcommand{\im}{\textrm{i}}

\newcommand{\wt}{\widetilde}

\begin{document}



\thispagestyle{empty}

\vspace*{100pt}
\begin{center}
\textbf{\Large{Conformal isometry of the six-dimensional black string and dualities of null geodesics}}\\
\vspace{50pt}
\large{David D. K. Chow}
\end{center}

\begin{center}
\textit{Max Planck Institute for Gravitational Physics (Albert Einstein Institute),\\
Am M\"{u}hlenberg 1, 14476 Potsdam-Golm, Germany}\\
{\tt david.chow@aei.mpg.de}\\
 \vspace{30pt}
 {\bf Abstract\\}
 \end{center}
We show that the extreme six-dimensional black string admits a conformal isometry under inversion.  Duality relations between null geodesics of various brane geometries are demonstrated, some of which have a geometrical origin through an optical metric.
\newpage


\section{Introduction}


The 4-dimensional extreme Reissner--Nordstr\"{o}m metric admits a conformal isometry that interchanges the horizon with asymptotic infinity \cite{coutor}.  Working in isotropic coordinates, this conformal isometry corresponds to an inversion of the radial coordinate $r$.  Although it is unclear whether the symmetry has a deeper meaning at the quantum level, it has received renewed attention \cite{Bizon:2012we} because of its connection to the classical instability of the solution under perturbations of a massless scalar field \cite{Aretakis:2012ei}.  The instability appears to be a general phenomenon, having been generalized to a wider class of solutions and fields of higher spin \cite{Lucietti:2012sf} and to higher dimensions \cite{Murata:2012ct}.

In contrast, although it is simple to invent metrics that satisfy a conformal isometry under inversion, examples of physical interest have rarely been identified.  One known class is 4-dimensional black holes: the extreme Reissner--Nordstr\"{o}m solution and generalizations with a cosmological constant \cite{Brannlund:2003gj} or with two electric charges, regarded as a solution supergravity \cite{Godazgar:2017igz}.  A second example is the D3-brane of 10-dimensional type IIB supergravity \cite{Gibbons:1993sv} (see also \cite{Gubser:1998kv, Gibbons:2011sg}).  In fact, \cite{Gibbons:1993sv} gives a family of geometries admitting an inversion conformal isometry, but only notes explicitly these two examples.  A purpose of this paper is to highlight a third example, the dyonic extreme 6-dimensional black string \cite{Duff:1995yh}, whose self-dual limit \cite{Duff:1993ye} is in fact included in \cite{Gibbons:1993sv}, and to discuss related properties of null geodesics in brane solutions.

Using a unified metric ansatz for branes, we recover these two basic examples of 4-dimensional black holes and the D3-brane, and provide the conditions for the inversion conformal isometry to hold within this ansatz.  In particular, we show that if we have a non-singular near-horizon geometry that is a direct product of an anti-de Sitter spacetime and a sphere, then there are three possibilities: a 4-dimensional black hole, a D3-brane, or a 6-dimensional black string.

We then further examine properties of null geodesics within these spacetimes, extending the discussion to include M2- and M5-branes.  The differential equations governing spatial projections of null geodesics are Newtonian central force problems whose potentials are powers of the radius.  Dualities for these theories have been known beginning with their study by Newton, most famously between the Kepler problem and the simple harmonic oscillator.  More generally, there is a duality between the force laws $r^a$ and $r^b$ when $(a + 3) (b + 3) = 4$, whose non-trivial integer solutions are $(a, b) = (1, -2), (-4, -7), (-5, -5)$.  The case $(a, b) = (1, -2)$, mentioned above, is a key part of Newton's \textit{Principia}, which also shows that the self-dual case $(a, b) = (-5, -5)$ admits circular orbits through the origin.  Whereas Newton demonstrated dualities of the orbits using Euclidean geometry, Bohlin \cite{bohlin} considered the equations of motion in the $(a, b) = (1, -2)$ case, and Kasner \cite{kasner}, independently and slightly earlier, considered general $(a, b)$.  We shall therefore refer to this as Kasner--Bohlin duality.  The duality has been rediscovered and related to the Maupertuis principle \cite{Collas:1981nc, hochnuro, arnvas}, applied to the Schr\"{o}dinger equation \cite{faure} (see also references within \cite{Grant:1993rh}), and popularized \cite{arnold, needham}.  We exhibit dualities between the null geodesics of different geometries, comparing with the known Kasner--Bohlin dualities of Schwarzschild black holes in 4, 5 and 7 dimensions \cite{Gibbons:2011rh}.  In some cases, the dualities can be understood from an optical 2-metric \cite{Casey:2012wu}.  Self-duality of the D3-brane is related to the $(a, b) = (-5, -5)$ duality, and M2/M5-brane duality is related to the $(a, b) = (-4, -7)$ duality.  We conclude with an illustration of the inversion symmetry for a scalar wave equation and further discussion.


\section{Metric ansatz}


We start with the metric ansatz
\begin{align}
\df s^2 & = \prod_{I = 1}^N H_I^{-\wt{d}/(d + \wt{d})} \eta_{\mu \nu} \, \df x^\mu \, \df x^\nu + \prod_{I = 1}^N H_I^{d / (d + \wt{d})} (\df r^2 + r^2 \, \df \Omega_{\wt{d} + 1}^2) , & H_I & = 1 + \fr{Q_I}{r^{\wt{d}}} ,
\label{ansatz}
\end{align}
where $d$ is the dimension of the brane worldvolume, $D = d + \wt{d} + 2$ is the total spacetime dimension, $\eta_{\mu \nu} \, \df x^\mu \, \df x^\nu$ is a $d$-dimensional Minkowski metric, $\df \Omega_{\wt{d} + 1}^2$ is the round metric on $S^{\wt{d} + 1}$, and $Q_I$ are constants, with $\mu, \nu = 0, \ldots, d - 1$ and $I = 1, \, \ldots , N$.  This metric ansatz includes many black brane solutions in supergravity, in which case the constants $Q_I$ correspond to electric or magnetic charges for $p$-form fields.  	Special cases include: the M2-brane, for $(N, d, \wt{d}) = (1, 3, 6)$; the M5-brane, for $(N, d, \wt{d}) = (1, 6, 3)$; and D$p$-branes, for $(N, d, \wt{d}) = (1, p + 1, 7 - p)$ and $0 \leq p \leq 6$.

It is convenient to introduce the optical metric.  A static spacetime can be written in canonical form as $\df s^2 = - N^2 \, \df t^2 + g_{i j} \, \df x^i \, \df x^j$, where $N$ and $g_{i j}$ depend only on the spatial coordinates $x^i$.  It is well-known that the spatial projections of its null geodesics are geodesics of the optical metric $\df s^2_\textrm{o} = N^{-2} g_{i j} \, \df x^i \, \df x^j$.  An isometry of the optical metric is equivalent to a conformal isometry of the full spacetime metric \cite{Abramowicz:2002qf}, with the conformal factor determined by the transformation of $N^2$.  In this case, the optical metric is
\be
\df s_\textrm{o}^2 = \sum_{\mu = 1}^{d - 1} (\df x^\mu)^2 + \prod_{I = 1}^N H_I \, (\df r^2 + r^2 \, \df \Omega_{\wt{d} + 1}^2) .
\ee
It is also useful to introduce the optical 2-metric \cite{Casey:2012wu}
\be
\df s_2^2 = \prod_{I = 1}^N H_I (\df r^2 + r^2 \, \df \phi^2) ,
\label{opt2}
\ee
which is the optical metric restricted to an equatorial plane of the transverse sphere and a fixed worldvolume point.

Let us initially assume that all charges are equal, i.e.\ $Q_I = Q$ for $I = 1, \ldots , N$.  There is an extreme horizon at $r = 0$, in some cases singular.  Under the coordinate transformation
\be
r = \fr{Q^{2/\wt{d}}}{\wt{r}} ,
\ee
the optical metric takes the form
\begin{align}
\df s_\textrm{o}^2 & = \sum_{\mu = 1}^{d - 1} (\df x^\mu)^2 + \bigg( \fr{Q}{\wt{r}^{\wt{d}}} \bigg) ^{4/\wt{d} - N} \wt{H}^N (\df \wt{r}^2 + \wt{r}^2 \, \df \Omega_{\wt{d} + 1}^2) , & \wt{H} & = 1 + \fr{Q}{\wt{r}^{\wt{d}}} .
\end{align}
Therefore, an isometry of the optical metric implies that $N \wt{d} = 4$.  There are three positive-integer solutions: $(N, \wt{d}) = (1, 4), \, (2, 2), \, (4, 1)$.

We can further specialize by considering solutions with a non-singular $\textrm{AdS}_{d + 1} \times S^{\wt{d} + 1}$ near-horizon geometry.  The near-horizon geometry of \eq{ansatz} is
\be
\df s^2 = \bigg( \fr{Q}{r^{\wt{d}}} \bigg) ^{- N \wt{d} / (d + \wt{d})} \eta_{\mu \nu} \, \df x^\mu \, \df x^\nu + \bigg( \fr{Q}{r^{\wt{d}}} \bigg) ^{N d/(d + \wt{d})} \big( \df r^2 + r^2 \, \df \Omega_{\wt{d} + 1}^2 \big) ,
\ee
which is $\textrm{AdS}_{d + 1} \times S^{\wt{d} + 1}$ when
\begin{align}
N \wt{d} & = 4 , & d & = \wt{d} .
\end{align}
The three solutions then correspond to: the D3-brane, for $(N, d, \wt{d}) = (1, 4, 4)$ \cite{Gibbons:1993sv}; the self-dual 6-dimensional dyonic string, for $(N, d, \wt{d}) = (2, 2, 2)$; and the 4-dimensional extreme Reissner--Nordstr\"{o}m black hole, for $(N, d, \wt{d}) = (4, 1, 1)$ \cite{coutor}.

Further possibilities come from relaxing the assumption that all charges are equal.  The 4-dimensional extreme Reissner--Nordstr\"{o}m black hole can be generalized to have 2 independent electric charges, regarded as a solution of $\mathcal{N} = 4$ supergravity or of $STU$ supergravity with pairwise equal electric charges \cite{Godazgar:2017igz}.  Similarly, we can generalize the self-dual 6-dimensional black string by taking independent electric and magnetic charges, as we now elaborate.


\section{Dyonic string}


The bosonic Lagrangian of minimal $\mathcal{N} = (2, 0)$, 6-dimensional supergravity coupled to a tensor multiplet is
\be
\mathcal{L} = R \star 1 - \tf{1}{2} \star \df \varphi \wedge \df \varphi - \tf{1}{2} \star H \wedge H ,
\ee
where $\varphi$ is the scalar dilaton and $H = \df B$ is the 3-form Kalb--Ramond field strength.  There is an extreme dyonic black string solution \cite{Duff:1995yh}, with independent electric and magnetic charges.  The special case of equal electric and magnetic charges is the self-dual solution of \cite{Duff:1993ye}, which has $H = \star H$.  The Einstein frame metric is
\begin{align}
\df s^2 & = \fr{- \df t^2 + \df x^2}{\sr{H_\textrm{e} H_\textrm{m}}} + \sr{H_\textrm{e} H_\textrm{m}} \, (\df r^2 + r^2 \, \df \Omega_3^2) , & H_\textrm{e} & = 1 + \fr{Q}{r^2} , & H_\textrm{m} & = 1 + \fr{P}{r^2} ,
\label{dyonicstring}
\end{align}
where $Q > 0$ and $P > 0$ are electric and magnetic charges respectively.  The matter fields are
\begin{align}
H & = \fr{2 Q}{H_\textrm{e}^2 r^3} \, \df t \wedge \df r \wedge \df x + 2 P \, \textrm{vol} (S^3) , & \varphi & = \fr{1}{\sr{2}} \log \bigg( \fr{H_\textrm{e}}{H_\textrm{m}} \bigg) ,
\end{align}
where $\textrm{vol}(S^3)$ is the volume-form of the round $S^3$.

Under the coordinate transformation
\begin{align}
r & = \fr{r_0^2}{\wt{r}} , & r_0 & = (Q P)^{1/4} ,
\label{inversion}
\end{align}
which represents inversion about the self-dual radius $r_0$, we have
\be
\df s^2 = \fr{r_0^2}{\wt{r}^2} \, \df \wt{s}^2 ,
\ee
where $\df \wt{s}^2$ is identical to the original metric \eq{dyonicstring}, but with $r$ replaced by $\wt{r}$.  Therefore, the inversion \eq{inversion} is a conformal isometry of the spacetime.  In the self-dual case $Q = P$, the solution falls within \cite{Gibbons:1993sv}, in their notation, $d = 10$, $p = 5$, $\gamma_x = 1$, $\gamma_r = - 1$.

The fixed point set of an isometry is a totally geodesic submanifold.  Analogously, the fixed point set of a conformal isometry is a photon surface, i.e.\ a surface such that null geodesics initially tangent to the surface remain in the surface.  It is the fixed point set of the corresponding isometry of the optical metric.  Therefore, there is a photon surface given by the self-dual radius $r = r_0$.


\section{Null geodesics and dualities}


The spherical symmetry of $S^3$ implies that we may choose coordinates that restrict geodesics to an equatorial plane with polar coordinates $(r, \phi)$.  Using standard manipulations, the unparameterized null geodesics satisfy
\begin{align}
\bigg( \fr{\df r}{\df \phi} \bigg) ^2 & = \fr{(r^2 + Q) (r^2 + P)}{b^2} - r^2 , & \fr{1}{b^2} & = \fr{E^2 - X^2}{h^2} ,
\label{6drphi}
\end{align}
where $E = \dot{t}/\sr{H_\textrm{e} H_\textrm{m}}$, $X = \dot{x}/\sr{H_\textrm{e} H_\textrm{m}}$ and $h = \sr{H_\textrm{e} H_\textrm{m}} r^2 \dot{\phi}$ are constants of motion, which determine the impact parameter $b$, and dots denote affine parameter derivatives.  This is invariant under the conformal isometry, as required.  There is an unstable circular photon orbit at $r = r_0$ with $b = \sr{Q} + \sr{P}$ and $X = 0$, and, more generally with $X \neq 0$, a helical orbit.  There are null geodesics that asymptote to the photon surface from the outside and inside, given respectively by
\begin{align}
r & = r_0 \coth \bigg( \fr{r_0}{\sr{Q} + \sr{P}} \phi \bigg) , & r & = r_0 \tanh \bigg( \fr{r_0}{\sr{Q} + \sr{P}} \phi \bigg) ,
\label{curves}
\end{align}
and related by inversion.  As $\phi \rightarrow 0$, the outside curve asymptotes to the straight line $r \sin \phi = \sr{Q} + \sr{P}$.  At $\phi = 0$, the inside curve reaches the horizon, where its osculating circle, which is dual to the straight line, is $r = [r_0^2/(\sr{Q} + \sr{P})] \sin \phi$.  General solutions of \eq{6drphi} can be expressed in terms of the Weierstrass elliptic function; see e.g.\ \cite{Hackmann:2008tu, Gibbons:2011rh}.

More generally, the ansatz \eq{ansatz} gives
\be
\bigg( \fr{\df r}{\df \phi} \bigg) ^2 = \fr{r^4}{b^2} \prod_{I = 1}^N H_I - r^2 .
\ee
These trajectories in the $(r, \phi)$-plane are geodesics of the optical 2-metric \eq{opt2}.  An isometry of the optical metric obtained by taking a power of $r$ gives an isometry of the optical 2-metric, but the converse does not hold.

For the D3-brane, we have
\be
\bigg( \fr{\df r}{\df \phi} \bigg) ^2 = \fr{r^4 + Q_\textrm{D3}}{b^2} - r^2 .
\label{D3}
\ee
Comparing with \eq{6drphi}, we see a correspondence, between null geodesics of the 6-dimensional black string and the D3-brane.  For example, we can identify
\begin{align}
Q_\textrm{D3} & = Q_\textrm{6d} P_\textrm{6d} , & b_\textrm{D3}^2 & = b_\textrm{6d}^2 - Q_\textrm{6d} - P_\textrm{6d} , & b_\textrm{6d} \phi_\textrm{D3} & = b_\textrm{D3} \phi_\textrm{6d} .
\end{align}
The null geodesics that asymptote to the photon surface of the six-dimensional string \eq{curves} are mapped to those that asymptote to the photon surface, $r = r_0$, of the D3-brane,
\begin{align}
r & = r_0 \coth (\phi/\sr{2}) , & r & = r_0 \tanh (\phi/\sr{2}) ,
\end{align}
with $Q_\textrm{D3} = r_0^4$ and $b_\textrm{D3} = \sr{2} r_0$.

It is worth comparing the duality under inversion with the Kasner--Bohlin duality \cite{bohlin, kasner} of null geodesics in the 5-dimensional Schwarzschild solution \cite{Gibbons:2011rh}.  One analogously derives
\be
\bigg( \fr{\df r}{\df \phi} \bigg) ^2 = \fr{r^4}{b^2} + 2 m - r^2 ,
\label{5d}
\ee
where $m$ is the mass of the black hole.  There is a duality invariance under inversion of $r$, which swaps the $r^4$ and $r^0$ terms.  This also swaps $m$, the mass of the background solution, and $1/b^2$, the inverse impact parameter squared of the light ray, up to numerical factors.  In contrast, the self-duality under inversion for the 6-dimensional string or D3-brane holds for a fixed background geometry and fixed impact parameter.  The duality between the 6-dimensional string and the D3-brane mixes the background geometry and impact parameter in a more complicated way.  The Newtonian analogue is self-duality of an $r^{-5}$ force law, and Newton showed that the theory admits a circular orbit through the origin, although it is not the osculating circle mentioned above.  Comparing \eq{D3} and \eq{5d} gives a further duality between null geodesics of the D3-brane and the 5-dimensional Schwarzschild solution:
\begin{align}
b_\textrm{D3} & = b_\textrm{5d} = b , & Q_\textrm{D3} & = 2 m_\textrm{5d} b^2 .
\end{align}

Another example of Kasner--Bohlin duality is between null geodesics in 4- and 7-dimensional Schwarzschild solutions \cite{Gibbons:2011rh}.  There is a similar duality for M5- and M2-branes, whose null geodesics respectively satisfy
\begin{align}
\bigg( \fr{\df r}{\df \phi} \bigg) ^2 & = \fr{r^4}{b^2} \bigg( 1 + \fr{Q_\textrm{M5}}{r^3} \bigg) - r^2 , & \bigg( \fr{\df r}{\df \phi} \bigg) ^2 & = \fr{r^4}{b^2} \bigg( 1 + \fr{Q_\textrm{M2}}{r^6} \bigg) - r^2 .
\end{align}
These are related by
\begin{align}
\fr{r_\textrm{M5}^3}{Q_\textrm{M5}} & = \fr{Q_\textrm{M2}}{r_\textrm{M2}^6} , & 2 \phi_\textrm{M5} & = \phi_\textrm{M2} , & Q_\textrm{M2}^{1/6} b_\textrm{M5} & = Q_\textrm{M5}^{1/3} b_\textrm{M2} .
\label{M52}
\end{align}
We may further understand the duality through the optical 2-metrics
\begin{align}
\df s_\textrm{M5, 2}^2 & = H_5 (\df r^2 + r^2 \, \df \phi^2), & \df s_\textrm{M2, 2}^2 & = H_2 (\df r^2 + r^2 \, \df \phi^2) .
\end{align}
Under the $(r, \phi)$ coordinate change \eq{M52}, the optical 2-metrics are the same up to a multiplicative constant,
\be
Q_\textrm{M2}^{1/3} \df s_\textrm{M5, 2}^2 = 4 Q_\textrm{M5}^{2/3}  \, \df s_\textrm{M2, 2}^2 .
\ee
and so give the same unparameterized geodesics.  A related observation about the $(t, r)$ part of the sgeometries was made in \cite{Gubser:1998kv}.  In contrast, the Kasner--Bohlin duality between null geodesics in 4- and 7-dimensional Schwarzschild solutions does not have projectively equivalent optical 2-metrics \cite{Casey:2012wu}.  Although the M2- and M5-branes are electromagnetic duals of each other, it is curious that their geometries exhibit a partial duality.  The Newtonian analogue is duality between $r^{-4}$ and $r^{-7}$ force laws.

An examination of the optical 2-metric \eq{opt2} for positive-integer $N$ and $\wt{d}$ shows that its only non-trivial dualities not discussed so far are between $(N, \wt{d}) = (3, 1)$ and $(N, \wt{d}) = (3, 2)$, and within the family $(N, \wt{d}) = (4, 1)$.  3-charge black holes in 5-dimensional $STU$ supergravity have $(N, d, \wt{d}) = (3, 1, 2)$.  4-charge black holes in 4-dimensional $STU$ supergravity have $(N, d, \wt{d}) = (4, 1, 1)$, with instead $N = 3$ when one electric charge vanishes, giving a naked singularity.


\section{Scalar field}


The inversion conformal isometry can be seen in the equation of a test scalar field in the 6-dimensional black string background.  For illustration, we consider the scalar wave equation, allowing for a coupling to curvature, $(\square + \xi R) \Phi = 0$.  The Ricci scalar is
\be
R = \fr{(Q - P)^2}{(H_\textrm{e} H_\textrm{m})^{5/2} r^6} .
\ee
In the self-dual case $Q = P$, the constant dilaton and conformal invariance of $\star H \wedge H$ imply through the field equations that $R = 0$.  We separate variables, taking $\Phi = \phi (r) Y(\theta_i) \expe{- \im \omega t}$, where $Y(\theta_i)$ is a spherical harmonic on $S^3$ satisfying $\nabla^2 Y = - \ell (\ell + 2) Y$.  The radial function $\phi(r)$ satisfies
\be
\fr{\df^2 \phi}{\df r^2} + \fr{3}{r} \fr{\df \phi}{\df r} + \bigg( \omega^2 H_\textrm{e} H_\textrm{m} - \fr{\ell (\ell + 2)}{r^2} + \fr{\xi R}{\sr{H_\textrm{e} H_\textrm{m}}} \bigg)\phi = 0 .
\ee
The invariance under inversion of the first two terms is the well-known Kelvin transformation on the Laplace equation in flat space.  The last three terms are individually invariant under inversion.  For a conformally coupled scalar, $\xi = 1/5$, conformal invariance implies invariance under inversion, but we see that the inversion symmetry also holds for arbitrary $\xi$.

The minimally coupled massless scalar wave equation in the dyonic string background involves the modified Mathieu equation \cite{Cvetic:1999fv}, like for the D3-brane \cite{Gubser:1998iu}.  The inversion symmetry simply corresponds to the Mathieu equation being even.  It is possible that the appearance of special functions is somehow related to the geometric simplicity or symmetry of the background geometry.

There is a similar duality between the massless scalar wave equation in M2- and M5-brane backgrounds that can be understood from the duality between the optical 2-metrics.  However, the restriction to an equatorial plane requires spherical symmetry, and so the duality for scalar fields only holds for the s-wave, i.e.\ the $\ell = 0$ partial wave, as used in \cite{Gubser:1998kv}.  See also \cite{Taylor:1998tk, Cvetic:1999nq} for examples considering the scalar wave equation.


\section{Conclusion}


We have highlighted examples of physically relevant spacetimes that admit an inversion conformal isometry.  Dualities between null geodesics of several different black brane spacetimes were demonstrated.  There are different types of dualities, and some examples can be explained by isometries of the optical metric or an optical 2-metric.  The dualities for M2-, D3- and M5-branes are particularly striking, since there are corresponding electromagnetic dualities, but it is not clear why electromagnetic and null geodesic dualities are related.  Furthermore, the orbits for a central force law $r^{-n}$, for positive integers $n$, are integrable in terms of elliptic functions (but not trigonometric functions) only for $n = 4, 5, 7$, corresponding precisely to these branes.  It would be interesting if the inversion symmetry has further applications to extreme black brane instabilities or in quantum theory.


\section*{Acknowledgements}


This work is supported by the Max Planck Society through the ``Gravitation and Black Hole Theory'' independent research group.  I would like to thank Yi Pang for helpful discussions.


\end{document}